\documentclass[12pt]{iopart}

\usepackage{iopams} 
\usepackage{bm,graphics,graphicx,epsfig,color}

\def\apj{{\em Astrophys.\ J.\ }}
\def\apjl{{\em Astrophys.\ J.\ Lett.\ }}
\def\prd{{\em Phys.\ Rev.\ D }}
\def\cqg{{\em Class.\ Quantum Grav.\ }}

\begin{document}

\title[Capture by a Kerr black hole]{Capture of non-relativistic particles in eccentric orbits by a Kerr black hole}

\author{Clifford M. Will$^{1,2}$}

\ead{cmw@physics.ufl.edu}
\address{
$^1$ Department of Physics,
University of Florida, Gainesville FL 32611, USA
 \\
$^2$ GReCO, Institut d'Astrophysique de Paris, UMR 7095-CNRS,
Universit\'e Pierre et Marie Curie, 98$^{bis}$ Bd. Arago, 75014 Paris, France
}

\begin{abstract}
We obtain approximate analytic expressions for the critical value of the total angular momentum of a non-relativistic test particle moving in the Kerr geometry, such that it will be captured by the black hole.   The expressions apply to arbitrary orbital inclinations, and are accurate over the entire range of angular momentum for the Kerr black hole.   The expressions can be easily implemented in N-body simulations of the evolution of star clusters around massive galactic black holes, where such captures play an important role.
\end{abstract}
\pacs{04.70.Bw, 98.62.Js}

\maketitle

\section{Introduction and summary}
\label{sec:intro}

The behavior of large numbers of stars orbiting around a central massive rotating black hole is a subject of great current interest (for recent reviews, see \cite{alexander05,genzel10}).   Recently, Merritt {\em et al.}\cite{mamw1}  used N-body simulations to construct and evolve a cluster of stars around a massive black hole such as the one at the center of our own galaxy, and studied the perturbations induced by those stars on the orbits of hypothetical stars that might be used to test the black hole no-hair theorems \cite{cmwnohair,lalehcliff}.  In separate work, Merritt {\em et al.} also  analyzed the rate of injection of stars onto highly eccentric orbits around a massive black hole \cite{mamw2}.  Such orbits would lead to an ``extreme mass-ratio inspiral" (EMRI) that would result in the emission of detectable gravitational radiation. 

Generally, such N-body simulations must be evolved for long periods of time in order to allow for all correlations and relaxation mechanisms to be established, so as to develop realistic estimates for rates of particular events, such as EMRI production.  
In the case of \cite{mamw2}, the timescale was 10 Myr.

The simulations of Merritt {\em et al.}~\cite{mamw1} included the Newtonian gravitational attraction between the stars and the black hole and between each other, as well as post-Newtonian effects associated with the massive, spinning central black hole, including the dragging of inertial frames and the quadrupole moment of the black hole's field.  The simulations of \cite{mamw2} restricted attention to a non rotating black hole, but included the capture of stars by the black hole.  A star-star encounter can place one star on an orbit with sufficiently small angular momentum that it can be captured directly by the black hole.   This is a non-trivial effect: in typical simulations carried out in \cite{mamw2}, out of 50 stars in the cluster, 17 were captured by $t = 2$ Myr, and 30 were captured by $t= 10$ Myr.  

In \cite{mamw2} the condition for a capture was simple: if the total angular momentum per unit mass $L$ of a star after an encounter with another star was smaller than the critical value $L_c = 4GM/c$, where $M$ is the mass of the black hole, the star was deemed to be captured, the mass of the black hole was augmented by the star's mass, and the hole was given a suitable recoil.   This value of $L_c$ corresponds to the well-known angular momentum of an unstable circular orbit in the Schwarzschild geometry with conserved relativistic energy per unit mass $E =1$.   Grossman {\em et al.}\cite{levin2012} call this the  innermost bound spherical orbit (IBSO).
The latter condition is appropriate here, because the stars are very far from the black hole in non-relativistic orbits when they encounter one another.  Accordingly, one can treat them as being barely bound ($E=1$) from the point of view of the black hole's strong gravity.

However, there is every reason to expect that the typical massive black hole at the center of a galaxy, including our own, will be rotating, perhaps even very rapidly rotating.  As a consequence, the capture of a star will not be isotropic: stars in prograde orbits will be less likely to be captured than stars in retrograde orbits.   For example, for $E=1$ stars with orbits restricted to the equatorial plane, the critical angular momentum per unit mass for capture is given by
\begin{equation}
L_c = \frac{GM}{c} \left ( 2 + 2 \sqrt{1\mp \tilde{a}} \right ) \,,
\end{equation}
where $\tilde{a}$ is the dimensionless Kerr parameter, related to the black hole's angular momentum $J$ by $\tilde{a} \equiv Jc/GM^2 \, (0 \le \tilde{a} \le 1)$. The upper(lower) sign corresponds to prograde(retrograde) orbits
(see, for example, \cite{bardeen73,young76}).   This anisotropic capture could have important consequences for the evolution of the cluster, imparting a net angular momentum to it, for instance, because of the preferential loss of stars moving in a retrograde manner relative to the black hole.  

For orbits out of the equatorial plane, the problem is much more complicated, despite the presence of the additional  ``Carter'' constant of motion, which makes the mathematical problem completely integrable.  Generally one must resort to numerical solutions to study the behavior of orbits near this critical point \cite{Glampedakis,levin}.

Instead, we have found an analytic expression for the critical angular momentum for arbitrary orbits that is approximate, but surprisingly accurate over almost the whole range of black hole spins.  It has the form
\begin{equation}
L_c = \frac{GM}{c} \left ( 2 + 2 \sqrt{1 - \tilde{a} \cos i - \frac{1}{8}\tilde{a}^2 \sin^2 i F(\tilde{a}, \cos i) }
 \right ) \,.
 \label{Lcrit0}
\end{equation}
Because the total angular momentum is no longer a conserved quantity, we {\em define} $L$ in terms of the Carter constant: $L \equiv \sqrt{C}$ (all quantities are suitably scaled so as to be independent of the mass of the star).  We also define $\cos i \equiv L_z/L$, where $L_z$ is the conserved $z$-component of angular momentum per unit mass.  In the Schwarzschild limit and the Newtonian limit,  $L$ {\em is} the total conserved angular momentum and $i$ is the orbital inclination, with $0 \le i \le \pi/2$ corresponding to prograde orbits ($L_z >0$) and $\pi/2 \le i \le \pi$ corresponding to retrograde orbits ($L_z < 0$).  The function $F(\tilde{a}, \cos i) = 1 + (\tilde{a}/2) \cos i + \dots  $  is a series expansion in $\tilde{a}$ shown below in Eq.\ (\ref{Fseries}).   With the series carried to $O(\tilde{a}^4)$, the solution (\ref{Lcrit0}) agrees with numerical solutions for the critical angular momentum to better than $0.5$ percent for $0 \le \tilde{a} \le 0.9$; and to better than $5$ percent for $0.9 \le \tilde{a} \le 0.99$.  For the special case of the extreme Kerr black hole ($\tilde{a} =1$), we obtain a separate, but very accurate analytic fit for $L_c$ as a function of $\cos i$.    Equation (\ref{Lcrit0}) can be easily implemented in an N-body code as a capture criterion.

The rest of this paper provides details.  In Sec.\ \ref{basic} we review the basic equations for motion in the Kerr geometry, and in Sec.\ \ref{critical} we obtain the critical value of $L$ for capture in the case of $E=1$.  Concluding remarks are made in Sec.\ \ref{conclude}.  Henceforth, we use units in which $G=c=1$.

\section{Basic equations for geodesic motion in the Kerr geometry}
\label{basic}

The Kerr metric is given in Boyer-Lindquist coordinates by
\begin{eqnarray}
ds^2 &=& - \biggl ( 1 - \frac{2Mr}{\rho^2} \biggr ) dt^2
 + \frac{\rho^2}{\Delta} dr^2 + \rho^2 d\theta^2
 - \frac{4Mra}{\rho^2} \sin^2 \theta dt d\phi
 \nonumber \\
 && \quad + \biggl ( r^2 + a^2 + \frac{2Mra^2 \sin^2 \theta}{\rho^2} \biggr ) \sin^2 \theta d\phi^2  \,,
\end{eqnarray}
where $M$ is the mass and $a$ is the Kerr parameter, related to the angular momentum $J$ by 
$a \equiv J/M$, with $|a| \le M$ required for the metric to describe a black hole; $\rho^2 = r^2 + a^2 \cos^2 \theta$, and 
$\Delta = r^2 + a^2 - 2Mr$.  We will assume throughout that $a$ is positive.

Timelike geodesics in this geometry admit four conserved quantities: energy per unit mass of the particle $E$, angular momentum per unit mass $L_z$, Carter constant  $C$, and the norm of the four velocity $u^\alpha = dx^\alpha/d\tau$, given by 
\numparts
\begin{eqnarray}
\qquad \quad E &=& - u_0 = -g_{00} u^0 - g_{0\phi} u^\phi \,,
\label{energy}
\\
\qquad \quad L_z &=&  u_\phi = g_{0\phi} u^0 + g_{\phi \phi} u^\phi \,,
\label{Lz}
\\
\qquad \quad C &=& \rho^4 \bigl ( u^\theta \bigr )^2 + \sin^{-2} \theta L_z^2 + 
a^2 \cos^2 \theta ( 1 - E^2 ) \,,
\label{carter}
\\
g_{\alpha \beta} u^\alpha u^\beta &=& -1 \,.
\label{norm}
\end{eqnarray}
\endnumparts
The version of the Carter constant used here has the property that, in the Schwarzschild limit ($a \to 0$), $C \to L^2$, where $L$ is the total conserved angular momentum.

Solving Eqs.\ (\ref{energy}) and (\ref{Lz}) for $u^0$ and $u^\phi$, and Eq.\ (\ref{carter}) for $u^\theta$, we obtain
\begin{eqnarray}
u^0 &=& \frac{E g_{\phi\phi} + L_z g_{0\phi}}{\Delta \sin^2 \theta} \,,
\nonumber \\
u^\phi &=& - \frac{E g_{0\phi} + L_z g_{00}}{\Delta \sin^2 \theta} \,,
\nonumber \\
u^\theta &=& \rho^{-2} \bigl [ C - \sin^{-2} \theta L_z^2 - a^2 \cos^2 \theta ( 1 - E^2 ) \bigr ]^{1/2}  \,.
\end{eqnarray}
Substituting these into Eq.\ (\ref{norm}) and solving for $u^r$ yields, after some simplification,
\begin{equation}
\bigl ( u^r \bigr )^2 = \left ( \frac{r^4}{\rho^4} \right ) V(r) \,,
\end{equation}
where
\begin{equation}
V(r) =  \biggl ( 1+ \frac{a^2}{r^2} +\frac{2Ma^2}{r^3}  \biggr )E^2 -\frac{\Delta}{r^2} \biggl ( 1 + \frac{C}{r^2} \biggr ) + \frac{a^2 L_z^2}{r^4} - \frac{4MaEL_z}{r^3} \,.
\end{equation}
This potential can then be used to study turning points of the radial motion, and will be the key ingredient in studying the capture of particles.

\section{Critical angular momentum for capture}
\label{critical}

We now want to find the critical values of the constants such that an
orbit of a given energy $E$, angular momentum $L_z$, and Carter constant $C$ will not be ``reflected'' back to large distances, but instead will continue immediately to smaller values of $r$ and be captured by the black hole.  The turning points of the orbit are given by the values of $r$ where $u^r=0$, thus where $V(r)=0$.  The critical values of  $E$, $C$ and  $L_z$ are those for which the potential has an extremum at that same point, that is where $d[(r^4/\rho^4)V(r)]/dr = 0$.  But since we are choosing a point where $V(r) = 0$, this is equivalent to $dV(r)/dr= 0$.  The chosen sign for $V(r)$ also dictates that this point should be a minimum of $V(r)$, that is that $d^2 V(r)/dr^2 > 0$, corresponding to an unstable extremum.  

Stars will be injected into orbits that could pass very close to the black hole via interactions with other stars far away from the black hole, in the non-relativistic regime.  After the encounter, the orbit of such a star can be characterized by its Newtonian semi-major axis $\bar{a}$ (not to be confused with the Kerr parameter $a$), which will be large compared to $M$, and thus the energy per unit mass will be given approximately by
$E  \approx (1 - M/2\bar{a} ) \approx 1$.   If this encounter can be regarded as the final significant perturbation of the orbit by other stars, this energy, along with the other constants of motion in the Kerr geometry ($L_z, \, C$) will be conserved for the rest of the orbit.  As a consequence, we shall focus attention on orbits  for which $E = 1$. 

In this case, linear combinations of
the conditions $V(r)=0$ and $dV(r)/dr = 0$ yield two quadratic equations for $r$.  These correspond to special cases of, for example, Eqs.\ (2.12) and (2.13) of \cite{bardeenpressteuk}, or Eqs.\ (A1) and (A2) of \cite{levin2012}.   Solving for $r$ from one equation and substituting back into the other equation yields an algebraic equation for the critical value of $C$ in terms of $L_z$, $M$ and $a$. 

In order to make the transition to spherical symmetry and the connection to the Newtonian portions of the orbits more natural, we will {\em define} the new constants of motion, $L > 0$ and $\cos i$, related to $C$ and $L_z$ by
\begin{eqnarray}
L^2 &\equiv& C \,,
\nonumber \\
\cos i &\equiv& L_z/L \,.
\end{eqnarray}
In both the spherical and Newtonian limits, $L$ and $i$ correspond to the total angular momentum and the orbital inclination, respectively.  In the strong-field regime close to the black hole, these quantities remain constants of the motion, but their interpretation is less transparent.

For orbits in the equatorial plane, $ L = |L_z| $, and $i = 0$ for prograde orbits or $i=\pi$ for retrograde orbits.  For general orbits, $i$ ranges from $0$ to $\pi/2$ for prograde orbits, and from $\pi/2$ to $\pi$ for retrograde orbits.  

Scaling out the factors of $M$ by defining the dimensionless variables $\tilde{L} = L /M$, $\tilde{a} = a/M $, we find that the algebraic equation for the critical value $\tilde{L}_c$ takes the form
\begin{eqnarray}
0 &=&  Q \bigl \{ (1- \tilde{a}^2 \sin^2 i)\tilde{L}_c^8
- 4\tilde{a}\cos i \tilde{L}_c^7
- 2[8 -\tilde{a}^2 (3+7 \sin^2 i) ] \tilde{L}_c^6 
\nonumber \\
&& \quad 
+ 4 \tilde{a} \cos i [24 - \tilde{a}^2 (1+  9 \sin^2 i) ] \tilde{L}_c^5 
\nonumber \\
&& \quad
-\tilde{a}^2 [240 - 192 \sin^2 i - \tilde{a}^2(1+18\sin^2 i -27 \sin^4 i) ] \tilde{L}_c^4
\nonumber \\
&& \quad
+ 64 \tilde{a}^3 \cos i (5 -2\sin^2 i) \tilde{L}_c^3
-48 \tilde{a}^4 (5 - 4\sin^2 i) \tilde{L}_c^2 
\nonumber \\
&& \quad
+ 96 \tilde{a}^5 \cos i \tilde{L}_c - 16 \tilde{a}^6 \bigr \} \,,
\label{condition}
\end{eqnarray}
where $Q =1/[\tilde{L}_c^2 (\tilde{L}_c^4-12 \tilde{L}_c^2 + 24\tilde{a}\tilde{L}_c \cos i - 12 \tilde{a}^2 )^2]$.  The critical radius corresponding to this unstable circular orbit is given by
\begin{equation}
\tilde{r}_c = \frac{\tilde{L}_c^2 \left [ (\tilde{L}_c - a \cos i )^2 - 8a^2 \sin^2 i \right ] }{ \tilde{L}_c^4 - 12 (\tilde{L}_c - a \cos i )^2  -12 a^2 \sin^2 i} \,.
\end{equation}

Considering first the case of an equatorial orbit, with $i=0$ or $i= \pi$, the condition becomes
\begin{equation}
0= \frac{(\tilde{L}_c^2 \mp 4\tilde{L}_c +4\tilde{a})(\tilde{L}_c^2 \pm 4\tilde{L}_c -4\tilde{a})(\tilde{L}_c \mp \tilde{a})^4}{\tilde{L}_c^2(\tilde{L}_c^4-12 \tilde{L}_c^2 \pm 24\tilde{a}\tilde{L}_c - 12 \tilde{a}^2 )^2} \,.
\end{equation}
The only solutions that correspond to values of $r$ outside the event horizon and to maxima of the potential $V(r)$ are 
\begin{eqnarray}
\tilde{L}_{c+} &=& 2 + 2 \sqrt{1-\tilde{a}} \,, \quad i = 0 \, ({\rm prograde}) \,,
\nonumber \\
\tilde{L}_{c-} &=& 2 +  2 \sqrt{1+\tilde{a}} \,, \quad i = \pi \, ({\rm retrograde}) \,.
\end{eqnarray}
The corresponding critical radius is given by
\begin{equation}
\tilde{r}_{c\pm} = (1 + \sqrt{1 \mp a})^2  = \tilde{L}_{c\pm} \mp a \,.
\end{equation}
For $\tilde{a}=0$, the solutions merge to $\tilde{L}_c = 4$ and $\tilde{r}_c =4$, corresponding to the well-known critical angular momentum in the Schwarzschild metric for the unstable circular orbit with $E=1$.  

\begin{figure}[t]
\begin{center}

\includegraphics[width=4in]{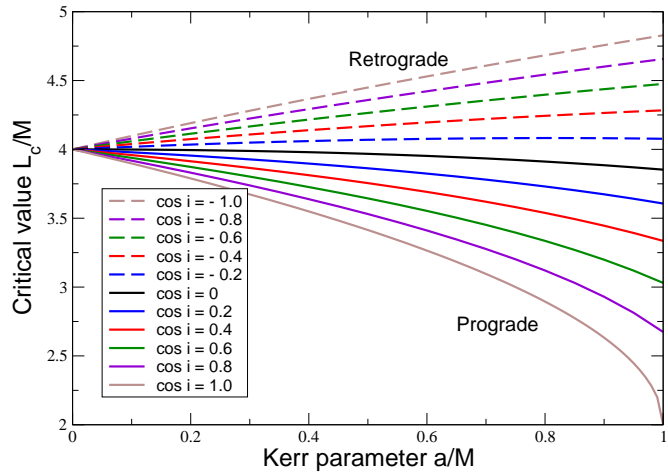}

\caption{\label{fig:critical} Critical angular momenta $L/M$ for capture vs. Kerr parameter $a/M$, for values of $\cos i $ ranging from $-1$ to $1$. }
\end{center}
\end{figure}

For arbitrary values of $i$, Eq.\ (\ref{condition}) must be solved numerically.  
The numerical solutions are plotted in Fig.\ 1, showing the critical value $L_c/M$ vs. $a/M$ for inclination angles ranging from $0$ to $\pi$.

Alternatively, we shall seek an analytic, albeit approximate solution.  To this end, we consider a solution for $\tilde{L}_c$  of the form
\begin{equation}
\tilde{L}_c = 2 + 2 \sqrt{1 - \tilde{a} \cos i - \frac{1}{8}\tilde{a}^2 \sin^2 i F(\tilde{a}, \cos i) } \,,
\label{Lcrit}
\end{equation}
which has the correct behavior both when $\tilde{a} = 0$ and when $i = 0$ or $i = \pi$ for arbitrary $a$.  We then solve Eq. (\ref{condition}) in a power series in $\tilde{a}$, obtaining
\begin{eqnarray}
F(\tilde{a}, \cos i) &=& 1 + \frac{1}{2} \tilde{a} \cos i + \frac{1}{64} \tilde{a}^2 (7 + 13 \cos^2 i) + \frac{1}{128}  \tilde{a}^3 \cos i (23 + 5 \cos^2 i)
\nonumber \\
&& + \frac{1}{2048} \tilde{a}^4 (55 + 340 \cos^2 i - 59 \cos^4 i) + O( \tilde{a}^5) \,.
\label{Fseries}
\end{eqnarray}
This series solution agrees with numerical solutions of Eq.\ (\ref{condition}) to better than $0.5$ percent for $0 \le \tilde{a} \le 0.9$; and to better than $5$ percent for $0.9 \le \tilde{a} \le 0.99$.  For $\tilde{a}=1$, the critical value can be fit to better than $0.5$ percent by the function
\begin{equation}
\tilde{L}_c = \cases{
   1.575 + 2.276 \sqrt{1 - 1.044 \cos i}   & : $-1 \le \cos i \le \sqrt{2/3}$ \\
   2/\cos i &: $ \sqrt{2/3} \le \cos i \le 1$  \,.\\} 
\end{equation}
The second solution is actually an exact solution for the range indicated.

Thus for a star in an orbit with $E=1$ and a given $\cos i = L_z/L$, plunge will occur for $L < M \tilde{L}_{c}$.
When the star is far from the black hole, the geometry is at the same time approximately Schwarzschild and approximately Newtonian.   In this case $L$ is the total angular momentum; it can be expressed in terms of the Newtonian semimajor axis $\bar{a}$ and eccentricity $e$ by 
$L^2  = M \bar{a} (1- e^2) = M r_p (1+e)$, where $r_p \equiv a(1-e)$ is the apparent pericenter distance inferred from the Newtonian orbit.   For large eccentricity ($e \approx 1$), then, capture will occur when 
\begin{equation}
r_p < r_{pc} \equiv  \frac{1}{2} M (\tilde{L}_{c\pm} )^2 \,.
\end{equation}
For Schwarzschild, the capture pericenter radius is $r_c = 8 M$; for extreme Kerr on the equatorial plane, the capture pericenter radius is $r_c = 2M (11.6 M)$ for prograde (retrograde) orbits.  

\section{Concluding remarks}
\label{conclude}

Implementation of this capture condition in an $N$-body code will depend on where along a given orbit the star's orbit elements are determined in order to find $r_p$, $i$ and $L_z$.  Presumably it is after the last important stellar encounter.  If the star is still relatively far from the black hole, then one can argue that relativistic corrections will be small compared to the variations shown in Eq.\ (\ref{Lcrit}).  If the star is close to the black hole when this evaluation is made, then it may be necessary to be more careful in applying Eq.\ (\ref{Lcrit}), for example by incorporating post-Newtonian corrections in the evaluation of the constants of motion for the orbit.  Given values of position and velocity of the star following the encounter, this can be done in a straightforward manner.  

It would be interesting to try to generalize the capture condition to value of $E$ less than unity, so as to handle more relativistic clusters.

\ack
This work was supported in part by the National Science Foundation,
Grant Nos.\ PHY 09--65133 \& 12--60995.  We thank the Institut d'Astrophysique de Paris for
its hospitality during the completion of this work.  We are grateful to 
Leor Barack,
Emanuele Berti, 
Kostas Glampedakis,
Shahar Hod,
Scott Hughes, 
Janna Levin, and
David Merritt
for useful comments.

\end{document}